\documentclass[aps,prd,preprint,nofootinbib]{revtex4}
\usepackage{mathrsfs}
\usepackage{amsfonts}
\usepackage{epsfig}
\usepackage{graphicx}
\newcommand{\bqa}{\begin{eqnarray}}
\newcommand{\eqa}{\end{eqnarray}}

\begin{document}
\title{The NLO QCD Corrections to $B_{c}$ Meson Production in
$Z^{0}$ Decays\\[9mm]}

\author{Cong-Feng Qiao$^{1,2} \footnote{qiaocf@gucas.ac.cn}$,
Li-Ping Sun$^{1}$\footnote{sunliping07@mails.gucas.ac.cn} and
Rui-Lin Zhu$^{1}$\footnote{zhuruilin09@mails.gucas.ac.cn}}
\affiliation{$^{1}$College of Physical
Sciences, Graduate University of Chinese Academy of Sciences \\
YuQuan Road 19A, Beijing 100049, China}
\affiliation{$^{2}$Theoretical Physics Center for Science Facilities
(TPCSF), CAS\\ YuQuan Road 19B, Beijing 100049, China}

\author{~\vspace{0.9cm}}

\begin{abstract}
\vspace{3mm} The decay width of $Z^{0}$ to $B_{c}$ meson is
evaluated at the next-to-leading order(NLO) accuracy in strong
interaction. Numerical calculation shows that the NLO correction to
this process is remarkable. The quantum chromodynamics(QCD)
renormalization scale dependence of the results is obviously
depressed, and hence the uncertainties lying in the leading order
calculation are reduced.

\vspace {7mm} \noindent {\bf PACS number(s):} 13.85.Ni, 14.40.Nd,
12.39.Jh, 12.38.Bx.

\end{abstract}

\maketitle
\section{Introduction}

$B_{c}$ meson has been attracting lots of attention in recent years.
As the sole heavy quark meson that contains two different heavy
flavors, its unique property attracts more and wide interests. Ever
since its first discovery at the TEVATRON \cite{CDF}, till now,
various investigations on its production and decays have been
carried out in aspects of theory \cite{Had1,Had2,Had3} and
experiment \cite{CDF,Ex1,Ex2}. $B_{c}$ meson provides an excellent
platform for testing the Standard Model(SM) and effective theories,
e.g., to see whether non-relativistic QCD(NRQCD) \cite{NRQCD} is
suitable for such system or not. In foreseeable near future, the
$B_c$ physics study at the Large Hadron Collider(LHC) will tell us
more about the nature of this special heavy bound system.

Of the $B_{c}$ meson production, apart from the direct ones, the
indirect yields, like in top \cite{Top} and $Z^{0}$ \cite{Z1,Z2,Z3}
decays, are also important sources. The process of $Z^{0}$ decays to
$B_{c}$ has an advantage of low background, but also with the
disadvantage of low production rate, which prompts the LEP-I
experiment unable to observe the $B_{c}$ signature \cite{Z1}. In the
future, if a high luminosity, $10^{34}cm^{-2}s^{-1}$ or higher,
electron-position collider, e.g. Internal Linear Collider(ILC)
\cite{ILC}, can set up, it would then be possible to study the $B_c$
meson indirect production in $Z^{0}$ decays. As estimated by
Ref.\cite{GigaZ}, there would be $10^{9}\sim10^{10}$ $Z^{0}$ events
produced each year at the ILC. Such kind of high luminosity
collider, the so-called ``Z factory'' \cite{Zfac}, will provide new
opportunities for both electroweak study and hadron physics.

The indirect production of $B_{c}$ meson in $Z^{0}$ decays was
evaluated by several groups at the leading order(LO) in strong
interaction \cite{Z1,Z2,Z3}. It is well-known that in charm- and
bottom-quark energy regions, the higher order corrections of strong
interaction are usually big, sometimes even huge. In order to make a
more solid prediction on the $B_c$ production in $Z^0$ decays, and
to depress the energy scale dependence lying at the LO calculation,
an evaluation on the next-to-leading order(NLO) correction is
necessary, which is the aim of this work.

The paper is organized as follows: after the Introduction, in
section II we repeat the leading order calculation on the $Z^0$ to
$B_c$ decay width. In section III, the NLO virtual and real QCD
corrections to Born level result are performed. In section IV, the
numerical calculation for the process at NLO accuracy is done. The
last section is remained for summary and conclusions.

\section{Calculation of The Born Level Decay Width}

\begin{figure}[t,m,u]
\centering
\includegraphics[width=15cm,height=10cm]{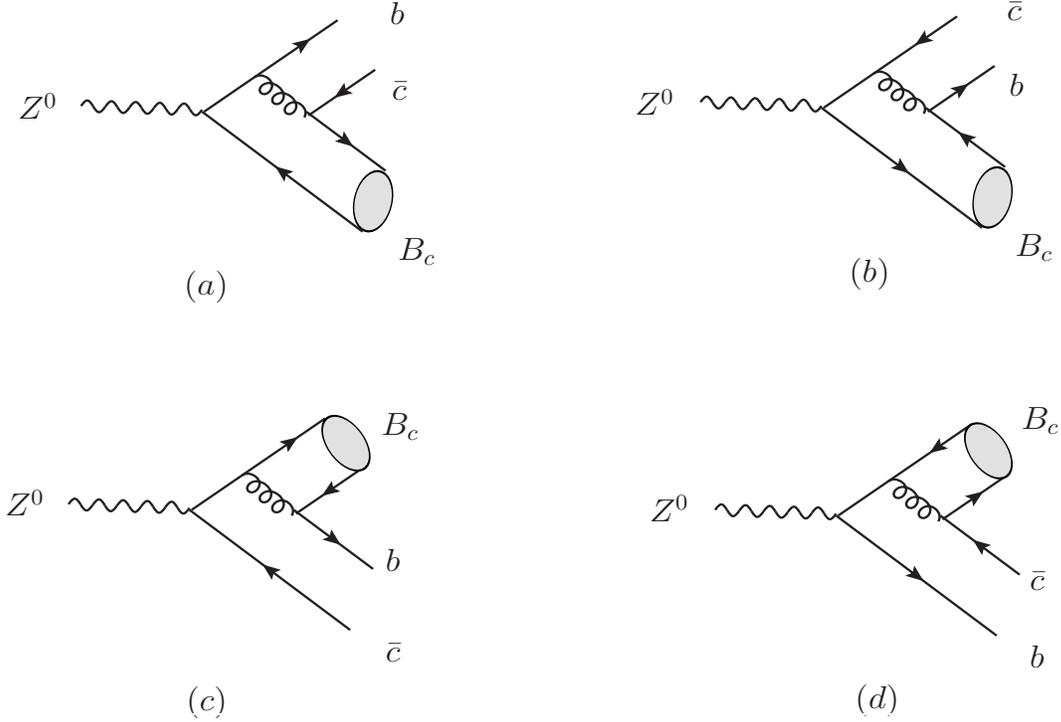}%
\caption{\small The leading order Feynman diagrams for ${B}_c$
production in $Z^{0}$ decays.} \label{graph1}
\end{figure}

At the leading order in $\alpha_{s}$, there are four Feynman
Diagrams for $B_{c}$ production in $Z^0$ decays, the
$Z^{0}(p_{1})\rightarrow {B}_c(p_{0})+\bar{c}(p_{5})+b(p_{6})$, as
shown in Fig.\ref{graph1}. The momentum of each particle is assigned
as: $p_{1}=p_{Z}$, $p_{3}=p_{\bar{b}}$, $p_{4}=p_{c}$,
$p_{5}=p_{\bar{c}}$, $p_{6} = p_{b}$, $p_{0} = p_{3} + p_{4}$, $p_3
= \frac{m_b}{m_c} p_4$. For $b$ and $c$ quark hadronization to
${B}_c$ meson, we employ the following commonly used projection
operator
\begin{eqnarray}
v(p_{3})\,\overline{u}(p_4)& \longrightarrow& {1\over 2 \sqrt{2}}
i\gamma_5(\not\!p_{0}+m_b+m_c)\, \times \left( {1\over
\sqrt{(m_b+m_c)/2}}~\psi_{{B}_c}(0)\right) \otimes \left( {{\bf
1}_c\over \sqrt{N_c}}\right)\label{eq:2}
\end{eqnarray}
Here, ${\bf 1}_c$ stands for the unit color matrix, and $N_c=3$ for
QCD. The nonperturbative parameter $\psi_{{B}_c}(0)$ is the
Schr\"{o}dinger wave function at the origin of the $\bar{b}c$ bound
states. In our calculation, the non-relativistic relation $m_{{B}_c}
= m_b+m_c$ is also adopted.
%
%

The LO amplitudes for ${B}_c$ production can then be readily
obtained with above preparations. They are:
\begin{eqnarray}
{\cal M}_{a}=\frac{\pi\alpha_{s}g\psi_{{B}_c}(0)C_{F}
\delta_{jk}}{6\sqrt{3m_{{B}_c}}\cos{\theta_{W}}}
~\bar{u}(p_{6})\gamma_{\mu}\frac{\not\!\epsilon(p_{1})
[(4\sin^{2}{\theta_{W}}-3)+3\gamma_{5}]}{(\not\!p_{1}
-\not\!p_{3}-m_{b})({p_{4}+p_{5}})^{2}}~i\gamma_{5}(
\not\!p_{0}+m_{{B}_c})\gamma^{\mu}v(p_{5})\;
, \label{eq:4}
\end{eqnarray}
\begin{eqnarray}
{\cal M}_{b}=\frac{\pi\alpha_{s}g\psi_{{B}_c}(0)C_{F}
\delta_{jk}}{6\sqrt{3m_{{B}_c}}\cos{\theta_{W}}}
~\bar{u}(p_{6})\gamma^{\mu}~i\gamma_{5}(\not\!p_{0}+m_{
{B}_c})\frac{\not\!\epsilon(p_{1})[(8\sin^{2}{\theta_{W}}
-3)+3\gamma_{5}]}{({p_{3}+p_{6}})^{2}(\not\!p_{1}-
\not\!p_{4}-m_{c})}\gamma_{\mu}v(p_{5})\;,\label{eq:5}
\end{eqnarray}
\begin{eqnarray}
{\cal M}_{c}=\frac{\pi\alpha_{s}g\psi_{{B}_c}(0)C_{F}
\delta_{jk}}{6\sqrt{3m_{{B}_c}}\cos{\theta_{W}}}
~\bar{u}(p_{6})\gamma^{\mu}~i\gamma_{5}(\not\!p_{0}+m_
{{B}_c})\gamma_{\mu}\frac{\not\!\epsilon(p_{1})[(8\sin
^{2}{\theta_{W}}-3)+3\gamma_{5}]}{(\not\!p_{1}-\not\!p
_{5}-m_{c})({p_{3}+p_{6}})^{2}}v(p_{5})\;,\label{eq:6}
\end{eqnarray}
and
\begin{eqnarray}
{\cal M}_{d}=\frac{\pi\alpha_{s}g\psi_{{B}_c}(0)C_{F}
\delta_{jk}}{6\sqrt{3m_{{B}_c}}\cos{\theta_{W}}}
~\bar{u}(p_{6})\frac{\not\!\epsilon(p_{1})[(4\sin^{2}
{\theta_{W}}-3)+3\gamma_{5}]}{({p_{4}+p_{5}})^{2}(\not
\!p_{1}-\not\!p_{6}-m_{b})}\gamma^{\mu}~i\gamma_{5}(\not
\!p_{0}+m_{{B}_c})\gamma_{\mu}v(p_{5})\; .\label{eq:6}
\end{eqnarray}
Here, $j$, $k$ are color indices, $C_F=4/3$ belongs to the $SU(3)$
color structure. $\theta_{W}$ is the Weinberg angle with the
numerical value $\sin^{2}{\theta_{W}}=0.23$.

The Born amplitude of the processes shown in Fig.\ref{graph1} is
then ${\cal M}_{Born} = {\cal M}_{a} + {\cal
M}_{b}+\mathcal{M}_{c}+\mathcal{M}_{d}$, and subsequently, the decay
width at LO reads:
\begin{eqnarray}
\mathrm{d}\Gamma_{Born}=\frac{1}{2m_{Z}}\frac{1}{3} \sum|{\cal
M}_{Born}|^{2}\mathrm{d}\textmd{PS}_{3}\; .\label{eq:6}
\end{eqnarray}
Here, $\sum$ symbolizes the sum over polarizations and colors of the
initial and final particles, $\frac{1}{3}$ comes from spin average
of initial $Z^{0}$ meson, $\mathrm{d}\textmd{PS}_{3}$ stands for the
integrals of three-body phase space, whose concrete form can be
expressed as:
\begin{eqnarray}
\mathrm{d}\textmd{PS}_{3}=\frac{1}{32\pi^3}\frac{1}{4m_Z^2}
\mathrm{d}s_{2}\mathrm{d}s_{1}\; ,\label{eq:7}
\end{eqnarray}
where $s_{1}=(p_{0}+p_{5})^{2}=(p_{1}-p_{6})^{2}$ and $s_{2}=
(p_{5}+p_{6})^{2}=(p_{1}-p_{0})^{2}$ are Mandelstam variables. The
upper and lower bounds of the above integration
are
\begin{eqnarray}
s_{1}^{max}=&&\frac{\sqrt{f[m_Z^2,s_2,m_{{B}_c}^2]\cdot
f[s_2,m_c^2,m^2_{b}]}+[m_Z^2-s_2-(m_b+m_c)^2](s_2+m_c^2
-m^2_{b})}{2s_2}\nonumber\\&& +m_{{B}_c}^2+m_c^2\;
,\label{eq:9}
\end{eqnarray}

\begin{eqnarray}
s_{1}^{min}=&&-\frac{\sqrt{f[m_Z^2,s_2,m_{{B}_c}^2]\cdot
f[s_2,m_c^2,m^2_{b}]}-[m_Z^2-s_2-(m_b+m_c)^2](s_2+m_c^2-
m^2_{b})}{2s_2}\nonumber\\&& +m_{{B}_c}^2+m_c^2 \label
{eq:10}\end{eqnarray} and

\begin{eqnarray} s_{2}^{max} =
[m_{Z}-(m_{b}+m_{c})]^{2}\; ,\;  s_{2}^{min} =(m_{c}+m_{b})^{2}
\end{eqnarray}

with

\begin{eqnarray}
f[x,y,z]&=&(x-y-z)^2-4yz\; .
\end{eqnarray}

\section{The Next-to-Leading Order Corrections}
\begin{figure}[t,m,u]
\centering
\includegraphics[width=15cm,height=6cm]{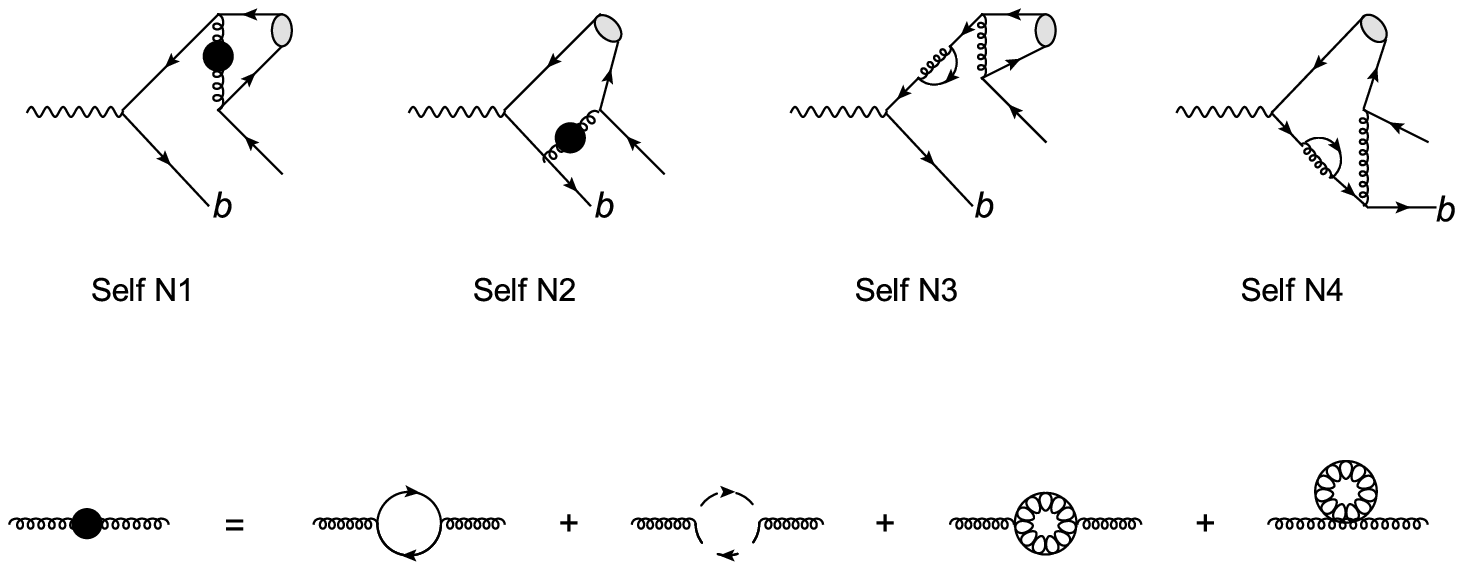}%
\caption{\small The self-energy diagrams in virtual corrections.}
\label{graph2}
\end{figure}

\begin{figure}[t,m,u]
\centering
\includegraphics[width=15cm,height=14cm]{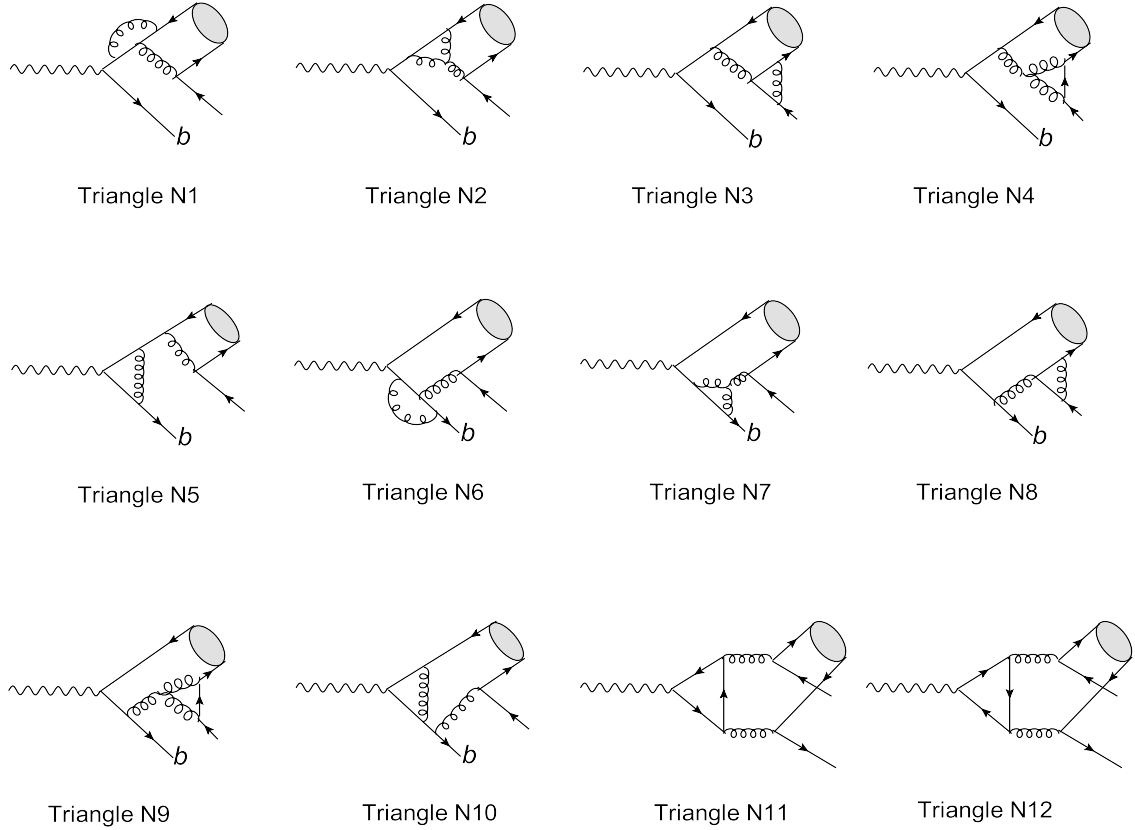}\vspace{-2.5cm}
\caption{\small The triangle diagrams in virtual corrections.}
\label{graph3}
\end{figure}

\begin{figure}[t,m,u]
\centering
\includegraphics[width=15cm,height=12cm]{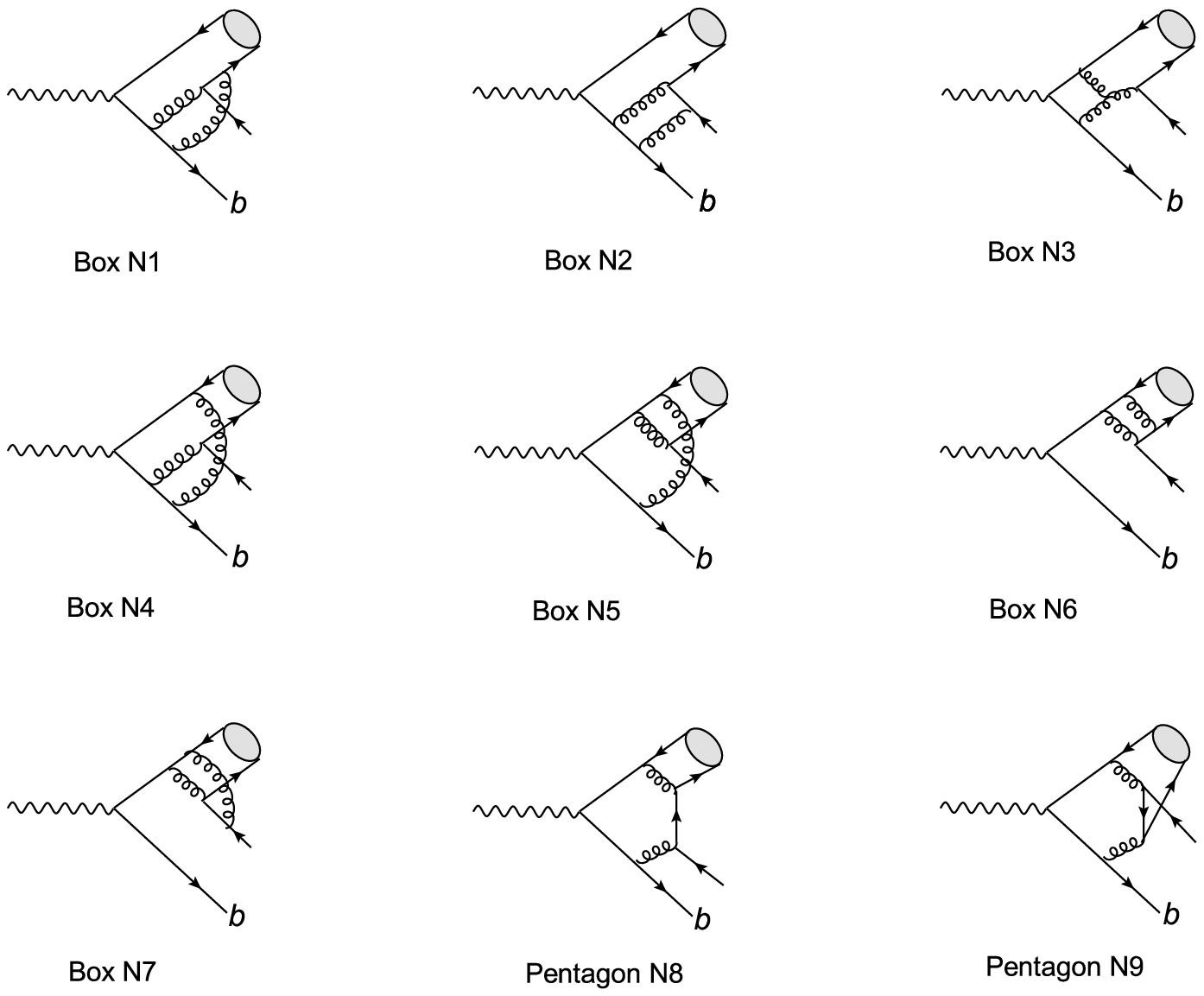}%
\caption{\small The box and pentagon diagrams in virtual
corrections.} \label{graph4}
\end{figure}

\begin{figure}[t,m,u]
\centering
\includegraphics[width=15cm,height=11cm]{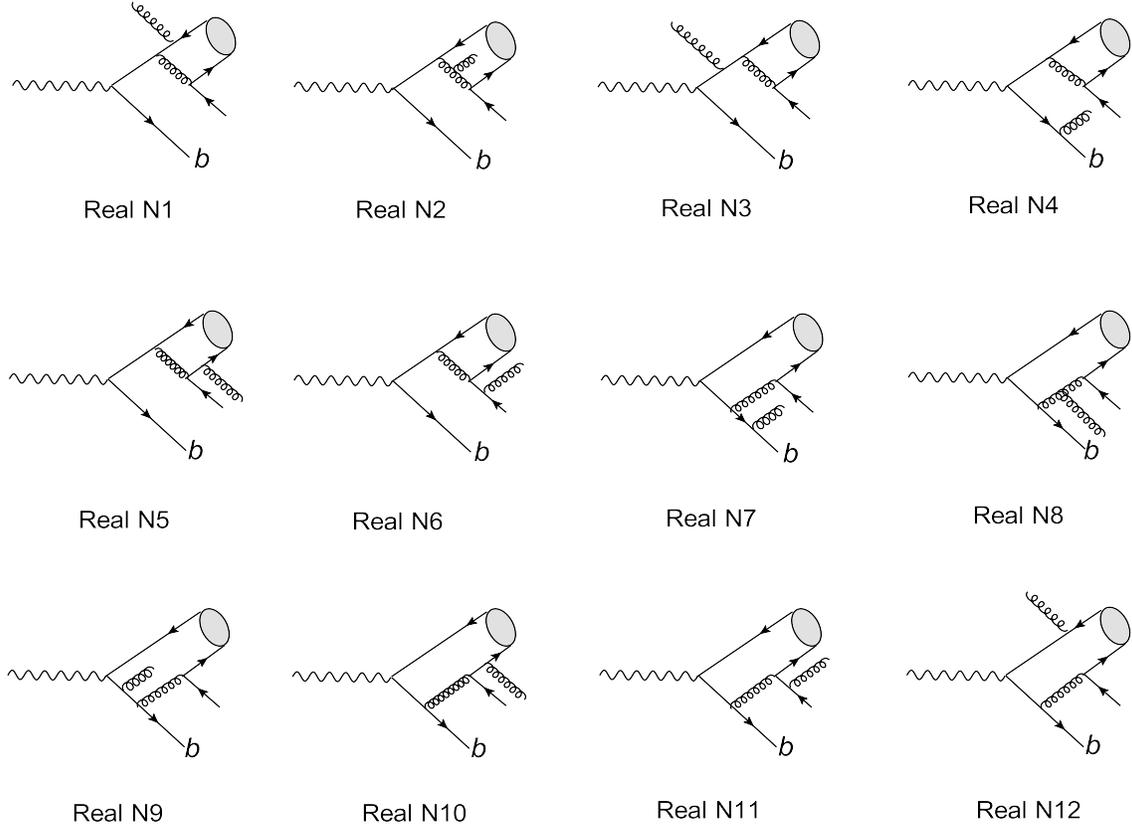}%
\caption{\small The real correction Feynman diagrams that contribute
to the production of $B_c$.} \label{graph5}
\end{figure}

At the next-to-leading order, the $Z^{0}$ boson decay to ${B}_c$
includes the virtual and real QCD corrections to the leading order
process. For the two kinds of vertices, $Z\rightarrow \bar{b}b$ and
$Z\rightarrow \bar{c}c$, we need only to consider one of them, e.g.
as shown in Figs.\ref{graph2}-\ref{graph5}, since they are similar.
For the virtual corrections, the decay width at the NLO can be
formulated as
\begin{eqnarray}
\mathrm{d}\Gamma_{Virtual}=\frac{1}{2m_{Z}}\frac{1}{3}
\sum2\textmd{Re} ({\cal M}_{Born}^{*} {\cal M}_{Virtual})
\mathrm{d}\textmd{PS}_{3}\; .\label{eq:12}
\end{eqnarray}

In virtual corrections, the ultraviolet(UV) and infrared(IR)
divergences exist universally. We use the dimensional regularization
scheme to regularize both UV and IR divergences, similar as
performed in Ref.\cite{DR}, and use the relative velocity $v$ to
regularize the Coulomb divergence \cite{vre}. According to the power
counting rule, the UV divergences exist merely in self-energy and
triangle diagrams, which can be canceled by counter terms. The
renormalization constants include $Z_{2}$, $Z_{3}$, $Z_{m}$, and
$Z_{g}$, corresponding to quark field, gluon field, quark mass, and
strong coupling constant $\alpha_{s}$, respectively. Here, in our
calculation the $Z_{g}$ is defined in the
modified-minimal-subtraction ($\mathrm{\overline{MS}}$) scheme,
while for the other three the on-shell ($\mathrm{OS}$) scheme is
adopted, which tells
\begin{eqnarray}
&&\hspace{-0.3cm}\delta Z_m^{OS}=-3C_F\frac{\alpha_s}{4\pi}
\left[\frac{1}{\epsilon_{UV}}-\gamma_{E}+\ln\frac{4\pi\mu^{2}}
{m^{2}}+\frac{4}{3}+{\mathcal{O}}(\epsilon)\right]\; ,\nonumber
\\ &&\hspace{-0.3cm}\delta Z_2^{OS}=-C_F\frac{\alpha_s}{4\pi}
\left[\frac{1}{\epsilon_{UV}}+\frac{2}{\epsilon_{IR}}-3\gamma_
{E}+3\ln\frac{4\pi\mu^{2}}{m^{2}}+4+{\mathcal{O}}(\epsilon)\right]
\; ,\nonumber\\
&&\hspace{-0.3cm}\delta Z_3^{OS}= \frac{\alpha_s}{4\pi}\left[(
\beta^{'}_0-2C_A)(\frac{1}{\epsilon_{UV}}-\frac{1}{\epsilon_{IR}})-
\frac{4}{3}T_{f}(\frac{1}{\epsilon_{UV}}-\gamma_{E}+\ln\frac{4\pi\mu^{2}}
{m_{c}^{2}})\right.\nonumber\\
&&~~~~~~~~~~~~~~~~~~~~~~~~~~~~~~~~~~~\left.-\frac{4}{3}T_{f}
(\frac{1}{\epsilon_{UV}}-\gamma_{E}+\ln\frac{4\pi\mu^{2}}{m_{b}^{2}})+
{\mathcal{O}}(\epsilon)\right]\; ,\nonumber\\
&&\hspace{-0.3cm}\delta Z_g^{\overline{MS}}=-\frac{\beta_0}{2}
\frac{\alpha_s}{4\pi}\left[\frac{1}{\epsilon_{UV}}-\gamma_{E}+
\ln4\pi+{\mathcal{O}}(\epsilon)\right]\; .\label{eq:13}
\end{eqnarray}
Here, the mass $m$ in $\delta Z_m^{OS}$ and $\delta Z_2^{OS}$
represents $m_{b}$ or $m_{c}$;
$\beta_{0}=(11/3)C_{A}-(4/3)T_{f}n_{f}$ is the one-loop coefficient
of the QCD beta function; $n_{f}=5$ is the number of active quarks
in our calculation; and $\beta_{0}' = (11/3)C_{A}-(4/3)T_{f}n_{lf}$
with $n_{lf}=3$ being the number of light-quark flavors; $C_{A}=3$
and $T_{F}=1/2$ attribute to the SU(3) group; $\mu$ is the
renormalization scale. Note, since the terms related to $\delta
\text{Z}_3^{OS}$ cancel with each other, the full NLO result is
independent of the renormalization scheme of the gluon field.

In dimensional regularization scheme, $\gamma_{5}$ is an object hard
to handle; especially in the process that contains the vector-axial
current, things become more complicated. In our work, we adopt the
scheme provided in Ref.\cite{gamma5}, where the following rules must
be obeyed:

I. the cyclicity is forbidden in traces involving odd number of
$\gamma_{5}$.

II. For the certain diagrams that contribute to a process, we must
write down the amplitudes starting at the same vertex, named the
reading point.

III. As a special case of rule II, if the anomalous axial current
exists, the reading point of the anomalous diagrams must be the
axial vector vertex, in order to guarantee the conservation of the
vector current.

By utilizing this rule in our process, the two anomalous diagrams
denoted as $\mathrm{Triangle N11}$ and $\mathrm{Triangle N12}$ in
Fig.\ref{graph3} are calculated, and the UV divergences in these two
diagrams are canceled by each other. To deal with the $\gamma_{5}s$
except for what in anomalous diagrams, the cyclicity is employed to
move the $\gamma_{5}s$ together and then are contracted by
$\gamma_{5}^{2}=1$. Hence, if a trace contains even number of
$\gamma_{5}$, there will be no $\gamma_{5}$ left. Otherwise, after
the contraction of odd number of $\gamma_{5}$, one remains.

In the virtual correction, IR divergences remain in the triangle and
box diagrams. Of all the triangle diagrams, only two have IR
divergences, which are denoted by $\mathrm{Triangle N3}$ and
$\mathrm{Triangle N8}$ in Fig.\ref{graph3}. Of the diagrams in
Fig.\ref{graph4}, $\mathrm{Box N3}$ has no IR divergence,
$\mathrm{Box N6}$ has merely the Coulomb singularity,
$\mathrm{Pentagon N8}$ has both a Coulomb singularity and ordinary
IR divergence, and the remaining other diagrams have only the
ordinary IR divergences. We find that the combinations of
$\mathrm{Box N1+Box N4}$, $\mathrm{Box N5+Pentagon N9+Triangle N8}$,
and $\mathrm{Box N7+Triangle N3}$ are IR finite, while the remaining
IR singularities in $\mathrm{Box N2}$ and $\mathrm{Box N8}$ are
canceled by the corresponding parts in real corrections. The Coulomb
singularities existing in $\mathrm{Box N6}$ and $\mathrm{Pentagon
N8}$ can be regularized by the relative velocity $v$. The
$\frac{1}{\epsilon}$ terms are renormalized by the counter terms of
external quarks which form the ${B}_c$, while the $\frac{1}{v}$ term
will be mapped onto the wave function of the concerned heavy meson.
In the end, the IR and Coulomb divergences in virtual corrections
can be expressed as
\begin{eqnarray}
\mathrm{d}\Gamma_{virtual} ^{IR,Coulomb}=\mathrm{d}\Gamma_{Born}
\frac{ 8 \alpha_s}{3\pi} \left[ \frac{\pi^2}{v}-\frac{1}{ \epsilon}
+\frac{(m_{Z}^{2}+2m_{b}m_{c}-2p_{1}\cdot p_{0})x_{s} \ln x_{s}}
{m_{b}m_{c}(1-x_{s}^{2})} \frac{1}{\epsilon} \right]\label{eq:14}
\end{eqnarray}
with $p_{1}=p_{Z}$, $p_{0}=p_{B_{c}}$ and $x_{s} =-
\frac{1-\sqrt{1-4m_{b}m_{c}/(4m_{b}m_{c}+m_{Z}^{2}-2p_1 \cdot
p_{0})}} {1 + \sqrt{1-4m_{b}m_{c}/(4m_{b}m_{c}+m_{Z}^{2}-2p_1 \cdot
p_{0})}}$. Here, in this work $\frac{1}{\epsilon}$ in fact
represents $\frac{1}{\epsilon}-\gamma_{E}+\ln(4\pi\mu^{2})$.

Of the concerned process, there are $12$ different Feynman diagrams
in real correction, as shown in Fig.\ref{graph5}. Among them,
$\mathrm{Real N2}$, $\mathrm{Real N3}$, $\mathrm{Real N8}$, and
$\mathrm{Real N9}$ are IR-finite, meanwhile the combinations of
$\mathrm{Real N1+ Real N5}$ and $\mathrm{Real N10+Real N12}$ exhibit
no IR singularities as well, due to the reason of gluon connecting
to either $\bar{b}$ or $c$ quark of the final ${B}_c$ meson. The
remaining diagrams, $\mathrm{Real N4}$, $\mathrm{Real N6}$,
$\mathrm{Real N7}$, and $\mathrm{Real N11}$ are not IR singularity
free. To regularize the IR divergence, we enforce a cut on the gluon
momentum, the $p_{7}$. The gluon with energy $p_{7}^{0} < \delta$ is
considered to be soft, while $p_{7}^{0} > \delta$ is thought to be
hard. The $\delta$ is a small quantity with energy-momentum unit. In
this way, the IR term of the decay width can then be written as:
\begin{eqnarray}
\mathrm{d}\Gamma_{Real}^{IR}=\frac{1}{2m_{Z}}\frac{1}{3} \sum|{\cal
M}_{Real}|^{2}\; \mathrm{d}\textmd{PS}_{4}\mid_{soft}\;,\label{eq:15}
\end{eqnarray}
where $\mathrm{d}\textmd{PS}_4$ is the four-body phase space
integrants for real correction. Under the condition of
$p_{7}^{0}<\delta$, in the Eikonal approximation we obtain
\begin{eqnarray}
\mathrm{d}\textmd{PS}_{4}\mid_{soft}=\mathrm{d}\textmd{PS}_{3}
\frac{d^{3}p_{7}}{(2\pi)^{3}2p_{7}^{0}}\mid_{p_{7}^{0}<\delta}\; .
\end{eqnarray}
In the small $\delta$ limit, the IR divergent terms in real
correction can therefore be expressed as
\begin{eqnarray}
\mathrm{d}\Gamma_{Real}^{IR}=\mathrm{d}\Gamma_{Born}\frac{8\alpha_s}
{3\pi}\left\{\left(\frac{1}{\epsilon}-\ln(\delta^2)\right)
\left[1-\frac{(m_{Z}^{2}+2m_{b}m_{c}-2p_{1}\cdot p_{0})x_{s} \ln
x_{s}}{m_{b}m_{c}(1-x_{s}^{2})}\right] + \rm{finite\;terms}\right\}
.~~~ \label{eq:17}
\end{eqnarray}
Here, the $\ln(\delta^2)$ involved terms will be canceled by the
$\delta$-dependent terms in the hard sector of real corrections.
Comparing (\ref{eq:17}) with (\ref{eq:14}), it is obvious that the
IR divergent terms in real and virtual corrections cancel with each
other. In the case of hard gluons in real correction, the decay
width reads
\begin{eqnarray}
\mathrm{d}\Gamma_{Real}^{hard}=\frac{1}{2m_{Z}}\frac{1}{3}\sum|
{\cal M}_{Real}|^{2}\; \mathrm{d} \textmd{PS}_{4}\mid_{hard}\;
.\label{eq:18}
\end{eqnarray}
In this case, the phase space $\mathrm{d}\textmd{PS}_{4}\mid_{hard}$
can be expressed as
\begin{eqnarray}
\int\mathrm{d}\textmd{PS}_{4}\mid_{hard}=\frac{2}{(4\pi)^{6}}&&\frac{
\sqrt{(sy+m_{c}^{2}-m_{b}^{2})^{2}-4sym_{c}^{2}}}{y}\int_{{p_{0}}
^{0}_{-}}^{{p_{0}}^{0}_{+}}\mathrm{d}{p_{0}}^{0}\int_{-1}^{1}
\mathrm{d}\cos\theta_{c}\int_{0}^{2\pi}\mathrm{d}\phi_{c}\nonumber\\
&&\times\left\{\int_{\delta}^{{p_{7}}^{0}_{-}}\mathrm{d}{p_{7}}^{0}
\int_{y_{-}}^{y_{+}}\mathrm{d}y+\int_{{p_{7}}^{0}_{-}}^{{p_{7}}^{0
}_{+}}\mathrm{d}{p_{7}}^{0}\int_{\frac{(m_{b}+m_{c})^{2}}{s}}^{y_
{+}}\mathrm{d}y\right\} \label{eq:20}
\end{eqnarray}
with
\begin{eqnarray}
&&{p_{0}}^{0}_{-}=m_{b}+m_{c}\; ,\\
&&{p_{0}}^{0}_{+}=\frac{\sqrt{s}}{2}\; , \\
&&{p_{7}}^{0}_{-} = \frac{s-2\sqrt{s}{p_{0}}^{0}}
{2\sqrt{s}-2{p_{0}}^{0} + 2
\sqrt{|\overrightarrow{{p_{0}}}|}}\; ,\\
&&{p_{7}}^{0}_{+}= \frac{s-2\sqrt{s}{p_{0}}^{0}}
{2\sqrt{s}-2{p_{0}}^{0} - 2
\sqrt{|\overrightarrow{{p_{0}}}|}}\; ,\\
&&y_{-}=\frac{1}{s}[(\sqrt{s}-{p_{0}}^{0}-{p_{7}}^{0})^{2}-|
\overrightarrow{{p_{0}}}|^{2}-({p_{7}}^{0})^{2}
  -2|\overrightarrow{p_{0}}|p_{7}^{0}]\; ,\\
&&y_{+}=\frac{1}{s}[(\sqrt{s}-{p_{0}}^{0}-{p_{7}}^{0})^{2}-|
\overrightarrow{{p_{0}}}|^{2}-({p_{7}}^{0})^{2}
  +2|\overrightarrow{p_{0}}|{p_{7}}^{0}]\; .
\end{eqnarray}
Here, $y$ is a dimensionless parameter defined as $y =
(p_1-p_0-p_7)^2/s$ with $\sqrt{s}=m_{Z}$, and
$|\overrightarrow{p_{0}}|=\sqrt{({p_0}^0)^{2}-m_{{B}_{c}}^{2}}\; .$
%
The sum of the soft and hard sectors gives the total contribution of
real corrections, i.e., $\Gamma_{Real} = \Gamma_{Real}^{IR} +
\Gamma_{Real}^{hard}$.

With the real and virtual corrections, we then obtain the total
decay width of $Z$ boson to ${B}_c$ at the NLO accuracy of QCD
\begin{eqnarray}
\Gamma_{total}=\Gamma_{Born}+\Gamma_{Virtual}+\Gamma_{Real}+
{\mathcal{O}}(\alpha_s^4)\; .\label{eq:22}
\end{eqnarray}
In above expression, the decay width is UV and IR finite. In our
calculation the FeynArts \cite{feynarts} was used to generate the
Feynman diagrams, the amplitudes were generated by the FeynCalc
\cite{feyncalc}, and the LoopTools \cite{looptools} was employed to
calculate the Passarino-Veltman integrations. The numerical
integrations of the phase space were performed by the MATHEMATICA.

\section{Numerical results}

To complete the numerical calculation, the following ordinarily
accepted input parameters are taken into account:
\begin{eqnarray}
m_{b}=4.9\; \textmd{GeV},\; m_{c}=1.5\; \textmd{GeV},\; m_{Z}=91\;
\textmd{GeV},\;m_{W}=80\; \textmd{GeV},\;
\end{eqnarray}
\begin{eqnarray}
G_F=1.1660\times 10^{-5}\; \mathrm{GeV}^{-2}, \;
g^{2}=\frac{8G_{F}m_{W}^{2}}{\sqrt{2}}=0.4221,\label{eq:23}
\end{eqnarray}
\begin{eqnarray}
\psi_{{B}_c}(0)=\frac{R(0)}
{\sqrt{4\pi}}=0.3616\;\textmd{GeV}^{\frac{3}{2}}.\;
\end{eqnarray}
Here, $R(0)$ is radial wave function at the origin of $B_{c}$ meson,
which is estimated via potential model \cite{pot} and $G_F$ is Fermi
constant in weak interaction. The one loop result of strong coupling
constant is taken into account in our calculation, i.e.
\begin{eqnarray}
\alpha_s(\mu)=\frac{4\pi}{(11-\frac{2}{3}n_f)\ln(\frac{\mu^2}
              {\Lambda_{QCD}^2})}\; .\label{eq:24}
\end{eqnarray}

With the above preparation, one can readily obtain the decay width
of $Z^{0}$ to $B_c$ meson in NLO accuracy of pQCD. In practice, the
renormalization scale $\mu$ may run from $2m_{b}$ to $m_{Z}/2$. At
$\mu=2m_{b}$ and then $\alpha_{s}(2m_{b})=0.189$ with
$\Lambda_{QCD}$ chosen to be $128~\mathrm{MeV}$, the LO and NLO
decay widths of $Z^{0}\rightarrow B_{c}+\bar{c}b$ process are
\begin{eqnarray}
\Gamma_{LO}(Z^{0}\rightarrow B_{c}+\bar{c}b)=72.31~\mathrm{keV}
\end{eqnarray}
and
\begin{eqnarray}
\Gamma_{NLO}(Z^{0}\rightarrow B_{c}+\bar{c}b)=78.45~\mathrm{keV},
\end{eqnarray}
respectively. And, at the scale $\mu=m_{Z}/2$ and then
$\alpha_{s}(m_{Z}/2)=0.140$, the corresponding results are
\begin{eqnarray}
\Gamma_{LO}(Z^{0}\rightarrow B_{c}+\bar{c}b)=39.43~\mathrm{keV}
\end{eqnarray}
and
\begin{eqnarray}
\Gamma_{NLO}(Z^{0}\rightarrow B_{c}+\bar{c}b)=62.53~\mathrm{keV}\; .
\end{eqnarray}

\begin{figure}
\centering
\includegraphics[width=0.480\textwidth]{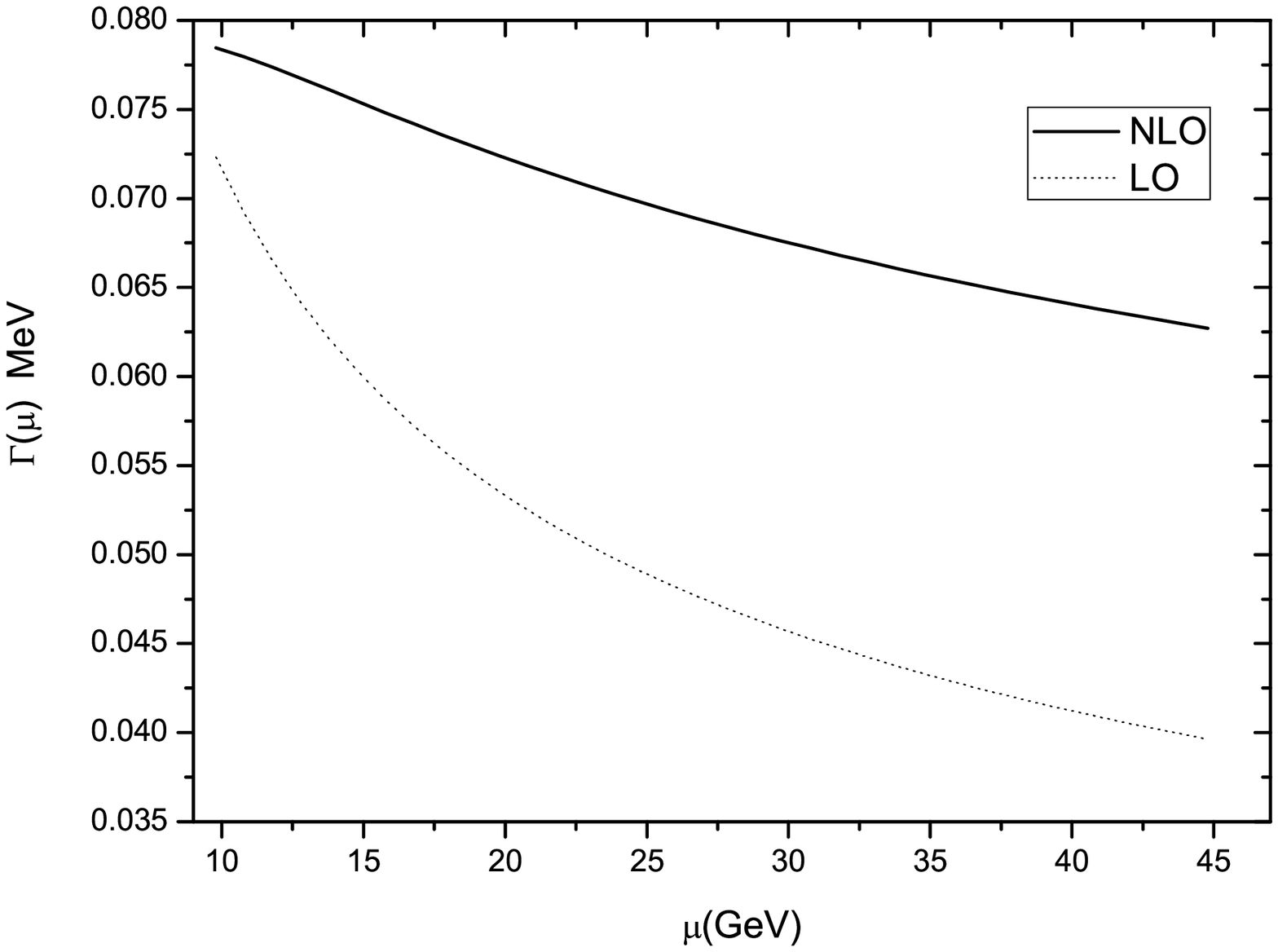}%
\hspace{5mm}
\includegraphics[width=0.480\textwidth]{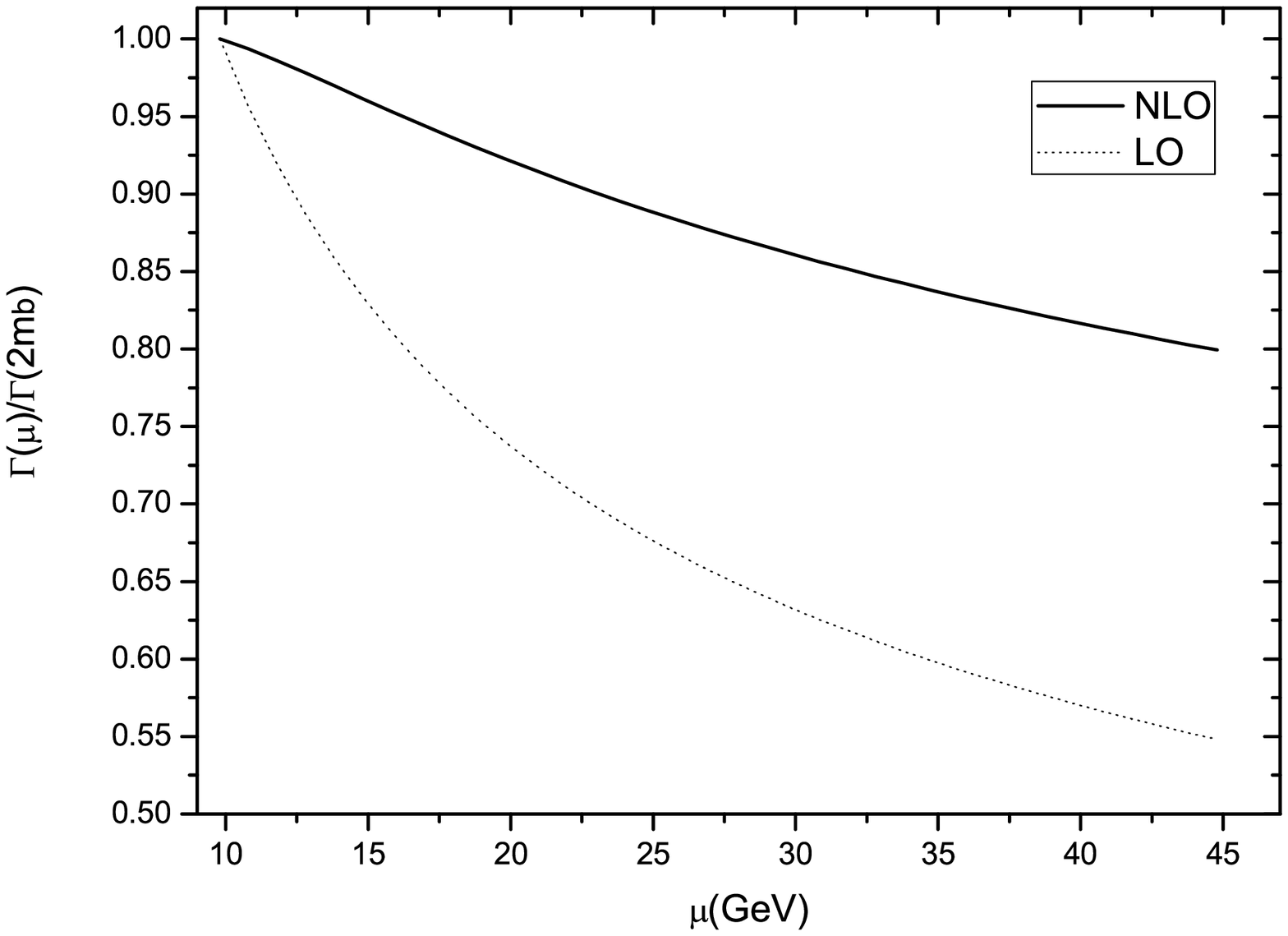}\hspace*{\fill}
\caption{\small The decay width $\Gamma(\mu)$(left) and the ratio
$\Gamma(\mu)/\Gamma(2m_c)$(right) versus renormalization scale $\mu$
in $Z^{0}$ boson decays.} \label{lpty1} \vspace{-0mm}
\end{figure}

Our LO result agrees with that existing in the literature
Ref.\cite{Z1} in case we take their inputs, i.e.
$\alpha_{s}(m_{Z}/2)=0.150$ and
$\psi_{{B}_c}(0)=0.369~\textmd{GeV}^{\frac{3}{2}}$. The above result
indicates that at high energy scale, the NLO QCD correction to the
decay width, or the $B_c$ production, is substantial. To see the
scale dependence of the LO and NLO results, the decay width
$\Gamma(\mu)$ and the ratio $\Gamma(\mu)/\Gamma(2m_{b})$  are shown
in Fig.\ref{lpty1} for $\mu$ varying from $2m_{b}$ to $m_{Z}/2$.
Calculation results indicate that after including the NLO QCD
corrections, as expected the energy scale dependence of the decay
width $\Gamma(Z^{0}\rightarrow B_{c} + \bar{c}b)$ is reduced
evidently.

\section{Summary and Conclusions}
In this work we have calculated the inclusive decay width of $Z^{0}$
boson to $B_{c}$ meson at the NLO accuracy of perturbative QCD.
Supposing that there will be copious $Z^{0}$ data in the future at
the ``Z-factory", our results are helpful to the precise study of
$B_{c}$ physics, and may also tell how well non-relativistic model
works for $B_{c}$ system.

Numerical results indicate that the NLO QCD correction slightly
increases the LO result for the process $Z^{0}\rightarrow
{B}_c+\bar{c}+b$ when $\mu$ is at the low energy scale of $2m_{b}$,
while it becomes huge, even comparable to the LO result, when $\mu$
runs to the scale of $m_{Z}/2$. We find that the energy scale
dependence of the decay width is depressed, as it should be, when
the next-to-leading order correction is taken into account, which
means the uncertainties in the theoretical estimation are reduced.

\vspace{.7cm} {\bf Acknowledgments}

This work was supported in part by the National Natural Science
Foundation of China(NSFC) and by the CAS Key Projects KJCX2-yw-N29
and H92A0200S2.



\newpage
\begin{thebibliography}{99}
\bibitem{CDF} CDF Collaboraten, F. Abe, {\em et al.}, Phys. Rev. Lett. {\bf 81},
 2432 (1998); Phys. Rev. D{\bf 58}, 112004 (1998).

\bibitem{Had1} K. Cheung, Phys. Lett. B{\bf 472}, 408 (2000); Chao-Hsi Chang,
 Yu-Qi Chen and R. J. Oakes, Phys. Rev. D{\bf 54}, 4344 (1996); Chao-Hsi Chang,
 Cong-Feng Qiao, Jian-Xiong Wang and Xing-Gang Wu, Phys. Rev. D{\bf 72},
 114009 (2005).

\bibitem{Had2} Chao-Hsi Chang and Yu-Qi Chen, Phys. Rev. D{\bf 48}, 4086
 (1993); Chao-Hsi Chang, Jian-Xong Wang and Xing-Gang Wu, Phys. Rev.
 D{\bf 70}, 114019 (2004); Chao-Hsi Chang, Yu-Qi Chen, Guo-Ping Han and
 Hung-Tao Jiang, Phys. Lett. B{\bf364}, 78 (1995)

\bibitem{Had3} K. Kolodziej, A. Leike and R. R\"uckl, Phys. Lett. B{\bf 355},
 337 (1995); Chao-Hsi Chang, Cong-Feng Qiao, Jian-Xiong Wang and Xing-Gang Wu,
 Phys. Rev. D{\bf 71}, 074012 (2005); Chao-Hsi Chang and Xing-Gang Wu, Eur.
 Phys. J. C{\bf 38}, 267 (2004).

\bibitem{Ex1} CDF Collaboration, D. Acosta {\em et al.}, Phys. Rev.
 Lett. {\bf 96}, 082002 (2006).

\bibitem{Ex2} CDF Collaboraten, F. Abe, {\em et al.}, Phys. Rev. Lett.
 {\bf 77}, 5176 (1996).

\bibitem{NRQCD} G. T. Bodwin, E. Braaten and G. P. Lepage, Phys. Rev. D{\bf 51},
 1125 (1995).

\bibitem{Top} Xing-Gang Wu, Phys. Lett. B{\bf 671}, 318 (2009); Peng-Sun,
 Li-Ping Sun and Cong-Feng Qiao, Phys. Rev. D{\bf 81}, 114035
 (2010); Chao-Hsi Chang, Jian-Xiong Wang and Xing-Gang Wu, Phys. Rev. D{\bf 77},
 014022 (2008); Cong-Feng Qiao, Chong-Sheng Li and Kuang-Ta Chao,
 Phys. Rev. D{\bf 54}, 5606 (1996).

\bibitem{Z1} Chao-Hsi Chang and Yu-Qi Chen, Phys. Rev. D{\bf 46}, 3845 (1992).

\bibitem{Z2} Zhi Yang, Xing-Gang Wu, Li-Cheng Deng, Jia-Wei Zhang and Gu Chen,
 Eur. Phys. J. C{\bf 71}, 1563 (2011); Li-Cheng Deng, Xing-Gang Wu, Zhi Yang,
 Zhen-Yun Fang and Qi-Li Liao, Eur. Phys. J. C{\bf 70}, 113 (2010).

\bibitem{Z3} V. V. Kiselev, A. K. Likhoded and M. V. Shevlyagin, Z. Phys. C{\bf 63},
 77 (1994); V. V. Kiselev, A. K. Likhoded and M. V. Shevlyagin, Phys. Atom.
 Nucl. {\bf 57}, 689 (1994), Yad. Fiz. {\bf 57}, 733 (1994).

\bibitem{ILC} G. Aarons et al., ILC collaboration, `International
 Linear Collider Reference Design Report Volume 2: PHYSICS AT THE
 ILC', arXiv:0709.1893[hep-ph].

\bibitem{GigaZ} J. Erler, {\it et al.}, Phys. Lett. B{\bf 486}, 125 (2000).

\bibitem{Zfac} Chao-Hsi Chang, Jian-Xiong Wang and Xing-Gang Wu,
 arXiv:1005.4723 [hep-ph].

\bibitem{DR} Yu-Jie Zhang, Ying-Jia Gao and Kuang-Ta Chao, Phys. Rev.
 Lett. {\bf 96}, 092001 (2006).

\bibitem{vre} M. Kr\"amer, Nucl. Phys. B{\bf 459}, 3 (1996).

\bibitem{gamma5} J. G. Korner, D. Kreimer and K. Schilcher, Z. Phys.
 C{\bf 54}, 503 (1992).

\bibitem{feynarts} T. Hahn, Comput. Phys. Commun. {\bf 140}, 418 (2001).

\bibitem{feyncalc} R. Mertig, M. B\"ohm and A. Denner, Comput.
 Phys. Commun. {\bf 4}, 345 (1991).

\bibitem{looptools} T. Hahn and M. Perez-Victoria, Comput. Phys. Commun.
 {\bf 118}, 153 (1999).

\bibitem{pot} A. Martin, Phys. Lett. B{\bf 93}, 338 (1980).

\end{thebibliography}
\end{document}